\documentclass[11pt,a4wide]{article}

\usepackage{amsfonts}
\usepackage{amsbsy} 
\usepackage{epsfig}
\usepackage{latexsym}
\input amssym.def
\input amssym.tex

\advance \topmargin by -\headheight
\advance \topmargin by -\headsep
\evensidemargin \oddsidemargin
\advance\hoffset by -3mm  
\advance\voffset by  8mm  
\def\bbbc{{\mathchoice {\setbox0=\hbox{$\displaystyle\rm C$}\hbox{\hbox
to0pt{\kern0.4\wd0\vrule height0.9\ht0\hss}\box0}}
{\setbox0=\hbox{$\textstyle\rm C$}\hbox{\hbox
to0pt{\kern0.4\wd0\vrule height0.9\ht0\hss}\box0}}
{\setbox0=\hbox{$\scriptstyle\rm C$}\hbox{\hbox
to0pt{\kern0.4\wd0\vrule height0.9\ht0\hss}\box0}}
{\setbox0=\hbox{$\scriptscriptstyle\rm C$}\hbox{\hbox
to0pt{\kern0.4\wd0\vrule height0.9\ht0\hss}\box0}}}}

\makeatletter
\@addtoreset{equation}{section}
\makeatother

\begin{document}
\hfuzz=100pt
\title{{\Large \bf{Classical Boundary-value Problem in Riemannian
Quantum Gravity and Taub-Bolt-anti-de Sitter Geometries}}}
\author{M M Akbar\footnote{Email:M.M.Akbar@damtp.cam.ac.uk}\\
Department of Applied Mathematics and Theoretical Physics,
\\ Centre for Mathematical Sciences,
\\ University of Cambridge, 
\\ Wilberforce Road,
\\ Cambridge CB3 0WA,
\\ U.K.}
\maketitle

\begin{center}
\date{DAMTP-2002-36}
\end{center}
\begin{abstract}
For an $SU(2)\times U(1)$-invariant $S^3$ boundary the classical
Dirichlet problem of Riemannian quantum gravity is studied for
positive-definite regular solutions of the Einstein equations with a
negative cosmological constant within biaxial Bianchi-IX metrics
containing bolts, i.e., within the family of Taub-Bolt-anti-de Sitter
(Taub-Bolt-AdS) metrics. Such metrics are obtained from the
two-parameter Taub-NUT-anti-de Sitter family. The condition of
regularity requires them to have only one free parameter ($L$) and
constrains $L$ to take values within a narrow range; the other
parameter is determined as a double-valued function of $L$ and hence
there is a bifurcation within the family.  We found that {\it{any}}
axially symmetric $S^3$-boundary can be filled in with at least one
solution coming from each of these two branches despite the severe
limit on the permissible values of $L$. The number of infilling
solutions can be one, three or five and they appear or disappear
catastrophically in pairs as the values of the two radii of $S^3$ are
varied. The solutions occur simultaneously in both branches and hence
the total number of independent infillings is two, six or ten.  We
further showed that when the two radii are of the same order and large
the number of solutions is two. In the isotropic limit this holds for
small radii as well. These results are to be contrasted with the
one-parameter self-dual Taub-NUT-AdS infilling solutions of the same
boundary-value problem studied previously.
\end{abstract}
\noindent
\section{Introduction}
When evaluating the path integral of Riemannian quantum gravity one often
requires an answer to the following Dirichlet problem: for a given closed
manifold $\Sigma$ with metric $h_{ij}$ what are the regular compact
solutions $({\cal{M}},g_{\mu\nu})$ such that $\partial{\cal{M}}=\Sigma$
and $g_{\mu\nu}|_{\partial{\cal{M}}}=h_{ij}$ and $g_{\mu\nu}$ satisfies
the Einstein field equations with appropriate matter fields or a
cosmological constant? Such classical solutions $({\cal{M}},g_{\mu\nu})$
are often referred to as infilling geometries of the `boundary'
$(\Sigma,h_{ij})$. They give semi-classical approximations to the path
integral and provide valuable insights into the nature of quantum gravity.

The general problem of finding all infilling solutions without any
restrictions on the possible forms of infilling metrics is a difficult
problem and often non-solvable. The boundary conditions and the
condition for regularity make the problem more non-trivial than just
finding Euclidean metrics solving Einstein's equations. Fortunately,
one often restricts oneself to rather symmetrical systems because of
physical or technical considerations.  For example, in cosmological
problems one usually assumes $(\Sigma,h_{ij})$ to be a homogeneous
manifold invariant under the transitive action of some Lie group $H$,
and assumes that a possible infilling 4-metric is of cohomogeneity one
under the group action of $H$. However, the generic orbits of the
infilling 4-metrics can be assumed to have less symmetry and invariant
under the group action of $H'\subseteq H$, provided $H'$ expands to
$H$ on the boundary $(\Sigma,h_{ij})$. In any case the assumption of
cohomogeneity (under the group action of $H$ or $H'$) reduces the
Dirichlet problem to a set of ordinary differential equations
(assuming, of course, no matter is present and only a cosmological
constant term exists) to be solved subject to the boundary data and
the condition of regularity in the interior. However, if complete
solutions of such cohomogeneity one metrics are known in advance, the
problem can be reduced to that of isometric embedding of a given
manifold $(\Sigma,h_{ij})$ in a manifold
$(\tilde{{\cal{M}}},g_{\mu\nu})$ one dimension higher. The infilling
geometry $({\cal{M}},g_{\mu\nu})$ is then the nonsingular compact
part(s) of the manifold $\tilde{{\cal{M}}}$ cut by the codimension-one
hypersurface $\Sigma$.  Consider the archetypal example of an $S^3$
boundary with the canonical round metric on it. For any value of the
(negative) cosmological constant, it can be filled in by parts of
$H^4$ (with the standard metric on it). On the other hand it is only
possible to embed the $S^3$ in a round $S^4$ if its radius is not
greater than the 4-radius of the $S^4$ (assuming a fixed positive
cosmological constant which determines the radius of $S^4$). Note that
in either of these infillings the generic orbits strictly have the
symmetry of the boundary. However, as mentioned above, one can in
principle consider infillings which are less symmetric in the
interior. For example, as shown in \cite{Akbar1} as part of a more
general boundary-value problem, any round $S^{3}$ can be filled in
with a regular biaxial Bianchi-IX solution with four-ball topology in
the presence of a negative cosmological constant.

In this paper we consider the boundary-value problem for an
$S^3$-boundary endowed with the following 3-metric
\begin{equation}
ds^2 = a^2({\sigma}_1^{~2}+{\sigma}_2^{~2})
              + b^2{\sigma}_3^{~2}, 
\end{equation}
where $\sigma_{i}$ are the left invariant one forms of $SU(2)$ and
hence the 3-boundary is invariant under the group-action of $SU(2)
\times U(1) \sim U(2)$. Such squashed $S^3$'s are known as Berger
spheres \cite{sak}. Following the discussion in the previous
paragraph, we naturally want to find solutions within the class of
cohomogeneity one metrics admitting $SU(2) \times U(1)$ action, i.e.,
within the family of positive definite biaxial Bianchi-IX Einstein
metrics. The general solution of such metrics is given by the
two-parameter Taub-NUT-(anti-)de Sitter family. When one imposes
regularity at the origin, one obtains two one-parameter families of
metrics: the self-dual Taub-NUT-(anti-)de Sitter and the
Taub-Bolt-(anti-)de Sitter. The former metric has a nut at the centre
and has a self-dual Weyl curvature tensor. The latter contains a bolt
-- a singular orbit -- corresponding to the two-dimensional
fixed-point set of the $U(1)$-action which is an $S^2$.

The Dirichlet problem within the self-dual Taub-NUT-AdS space was
studied in \cite{Akbar1}. This lead to a constraint on the possible
values of the two radii $(a,b)$ of the Berger sphere in the form of a
non-linear inequality between them.\footnote{In order to avoid
confusion due to divergent conventions in the literature, we will
reserve Taub-NUT-(anti-)de Sitter for the the whole two parameter
family and would call the one-parameter family containing (regular)
``nuts'' as self-dual Taub-NUT-(anti-)de Sitter family as the latter
has a self-dual Weyl tensor.} In this paper the problem will be
studied for the Taub-Bolt-AdS family. As we will see in Section 2.2
the condition of regularity of the bolt constrains the free parameter
($L$) to take values within a very narrow range on the real line with
a two-fold degeneracy in the other parameter ($M$) which is otherwise
determined by $L$. This is a rather stringent constraint compared to
the self-dual Taub-NUT-(anti-)de Sitter for which there is no such
{\it{a priori}} restriction on the values of the free parameter for
regularity. In this case, therefore, one would naturally expect some
stronger restrictions on the two radii of the Berger sphere for it to
qualify for a Taub-Bolt-AdS infilling. However, this is not the case
as we will see below.

Following the AdS/CFT correspondence \cite{Maldacena} the
Taub-Bolt-AdS and self-dual Taub-NUT-AdS spaces have recently garnered
much attention \cite{CEJM,Clarkson1,Emparan1,
Ghezelbash1,HHP,Mann1}. In particular it was shown in \cite{CEJM,HHP}
that a thermal phase transition occurs taking self-dual Taub-NUT-AdS to
Taub-Bolt-AdS in much the same way the Hawking-Page phase transition
occurs from hot AdS space to Schwarzschild-AdS \cite{HP}. In studying
the semi-classical transition amplitude of the process the action is
first calculated for a common finite boundary which encloses the
compact parts of the self-dual Taub-NUT-AdS and the Taub-Bolt-AdS
space such that they induce the same metric on this common
boundary. This common boundary is then taken to infinity to give the
transition amplitude from the complete self-dual Taub-NUT-AdS to the
complete Taub-Bolt-AdS spaces. Since the boundary is ultimately taken
to infinity, the two infilling spaces for the finite boundary need not
induce exactly the same metric on the boundary -- an approximate
matching is sufficient if that approximation is later justified when
the boundary is taken to infinity. In this paper we demonstrate that a
finite boundary formulation of the Dirichlet problem for the
Taub-Bolt-AdS space is in fact highly non-trivial. This has not been
addressed rigorously before.\footnote{Some investigations of the
similar boundary-value problem with the Taub-Bolt-de Sitter metric
were made in the context of quantum cosmology in \cite{Louko}.} Among
other results, we find that a typical axially symmetric $S^3$ boundary
admits multiple Taub-Bolt-AdS solutions. One can compare the situation
with Schwarzschild-AdS in a finite isothermal cavity in which case the
boundary is an $S^1 \times S^2$. Apart from the periodically
identified AdS space there are either two or zero AdS-Schwarzschild
infilling solutions depending on the boundary data \cite{BCM}. For a
biaxial $S^{3}$ boundary, which is a non-trivial $S^1$ bundle over
$S^2$, we find that the number of independent Taub-Bolt-AdS solutions
can be as high as ten. Finding explicit solutions and their possible
implications are left for future work.
\section{Positive-definite biaxial Binachi-IX Einstein metrics}
The Euclidean family of Taub-NUT-(anti-)de Sitter metrics is given by \cite{Carter,EGH,GP1}:
\begin{equation}
ds^2=\frac{\rho^{2} - L^{2}}{\Delta} d\rho^{2}+ \frac{4
L^{2}\Delta}{\rho^{2}-L^{2}}(d\psi+\cos \theta
d\phi)^{2}+(\rho^{2}-L^{2})(d\theta^{2}+\sin^{2}\theta d\phi^{2}) \label{met}
\end{equation}
where
\begin{equation}
\Delta=\rho^{2}-2M\rho +L^{2}+\Lambda( L^{4}+2
L^{2}\rho^{2}-\frac{1}{3}\rho^{4}).
\end{equation}
Here $L$ and $M$ are the two parameters and $0\leq\theta\leq\pi$,
$0\leq\phi\leq2\pi$, $0\leq\psi\leq4\pi/k$ ($k \in {\Bbb Z}$). When $k=1$, the surfaces of constant $\rho$ are topologically
$S^{3}$. These metrics are Einstein, i.e., they satisfy the Einstein
equation with a cosmological constant, $R_{\mu \nu}= \Lambda g_{\mu
\nu}$. They are axially symmetric, of
Bianchi type IX, i.e., of the form $ds^2 = N(r)^{2}dr^2 +
a^2(r)({\sigma}_1^{~2}+{\sigma}_2^{~2})+ b^2(r){\sigma}_3^{~2}$, where
$\sigma_{i}$ are left-invariant one forms on $S^{3}$ and hence are
invariant under the group action of $SU(2) \times U(1)$.
In the literature there has been considerable interest in biaxial
Bianchi type IX metrics for both Lorentzian and Riemannian signatures.

The general form (\ref{met}) is only valid for a coordinate patch for
which $\Delta \ne 0$. Because it is quartic, $\Delta (\rho)$ will have
four roots -- the so called ``bolts'' (two-spheres of constant radii)
of the metric (\ref{met}) which are the invariant points of the
Killing vector $\partial/ \partial \psi$.  However, if $\Delta$ has a
root at $\rho=|L|$, the fixed points there are zero dimensional and
such fixed points are called ``nuts'' \cite{GH}.  In general, bolts
and nuts are not regular points of the metric. The metric can be made
regular at a bolt or a nut provided it closes smoothly. For a bolt this is
possible if the following condition is satisfied \cite{Page}
\begin{equation}
\frac{d}{d\rho}\left(\frac{\Delta}{\rho^2-L^2} \right)_{(\rho=\rho_{bolt})}=\frac{1}{2kL}\label{cond},
\end{equation}
which amounts to imposing a relation between $L$ and $M$.  This is
also a necessary condition for a regular nut though it is not
sufficient to guarantee regularity. (Near a nut regularity requires
the metric to approach the flat metric on ${\Bbb E}^{4}$ which is not
guaranteed {\it{a priori}} by Eq.(\ref{cond}).)
\subsection{Self-dual Taub-NUT-(anti-)de Sitter}
A zero-dimensional fixed point-set of the $U(1)$-action occurs at
$\rho=|L|$, if $\Delta=0$, i.e., if
\begin{equation}
M= \pm L(1+\frac{4}{3}\Lambda L^{2}),
\end{equation}
which is precisely the condition for (anti-)self-duality of Weyl tensor
of the metric (\ref{met}), as one can check. It is easy to check that
for $k=1$ the metric approaches the flat metric on ${\Bbb E}^{4}$
guaranteeing that the metric at the nut is smooth satisfying
(\ref{cond}) trivially. The metric is then well-defined for the range
of $\rho$ starting from the nut until it encounters another zero of
$\Delta$.  For more details see \cite{Akbar1}. Assuming $L$ positive
and choosing the positive sign for $M$, $\Delta$ simplifies:
\begin{equation}
\Delta=(\rho - L)^{2}-\frac{1}{3}\Lambda(\rho+3 L)(\rho-L)^{3}.
\end{equation}
For the
purpose of the present paper, it is important to note that the condition of regularity
(\ref{cond}) has determined $M$ in terms of $L$ uniquely (modulo
sign) reducing the
resulting self-dual Taub-NUT-(anti-)de Sitter metrics effectively to
a one-parameter family.
\subsection{Taub-Bolt-(anti-)de Sitter}
For the purpose of
exposition consider the $\Lambda >0$ case first. Without any loss of
generality assume $\rho$ and $L$
are positive. Since a bolt occurs at a zero of $\Delta$, one can find $M$ in the
following way. Suppose the bolt occurs at $\rho=\rho_{bolt}$, then
\begin{equation}
M=\frac{1}{6}\,{\frac {3\,{\rho_{bolt}}^{2}-\Lambda\,{\rho_{bolt}}^{4}+3\,{L}^{2}+3\,\Lambda
\,{L}^{4}+6\,\Lambda\,{L}^{2}{\rho_{bolt}}^{2}}{\rho_{bolt}}}. \label{Mpos}
\end{equation}
By definition $\rho_{bolt}>L$. The regularity
condition (\ref{cond}) then gives (for $k=1$)
\begin{equation}
{\frac {-\Lambda\,{\rho_{bolt}}^{2}+{L}^{2}\Lambda+1}{\rho_{bolt}}}=\frac{1}{2L}.
\end{equation}
This is quadratic in $\rho_{bolt}$ and can have only one positive
root. It is easy to verify that this root will always occur within the
open interval $(L,\infty)$. Corresponding to this value of
$\rho_{bolt}$, $M$ is determined uniquely by (\ref{Mpos}). Thus, for
$\Lambda>0$ the bolt can be made regular for {\emph{any}} value of $L$
provided one fixes $M$ accordingly, i.e., through
(\ref{Mpos}). Therefore the imposition of condition (\ref{cond}) has
reduced the family to a truly one-parameter family.

In the case of negative cosmological constant the arguments are
slightly subtle. Denoting $-\Lambda=\lambda(>0)$ 
\begin{equation}
M=\frac{1}{6}\,{\frac {3\,{\rho_{bolt}}^{2}+\lambda\,{\rho_{bolt}}^{4}+3\,{L}^{2}-3\,\lambda
\,{L}^{4}-6\,\lambda\,{L}^{2}{\rho_{bolt}}^{2}}{\rho_{bolt}}}.
\end{equation}
The  regularity condition (\ref{cond}) (for $k=1$) then gives:
\begin{equation}
{\frac
{\lambda\,{\rho_{bolt}}^{2}-{L}^{2}\lambda+1}{\rho_{bolt}}}=\frac{1}{2L} \label{locabo}
\end{equation}
which can be rearranged into the more illuminating form:
\begin{equation}
2\,L\lambda\,\left({\rho_{bolt}}^{2}-L^2\right)-\left(\rho_{bolt}-2L\right)=0.\label{locabo1}
\end{equation}
Since $\rho_{bolt}>L$, this clearly requires $\rho_{bolt}>2L$. This
equation can be
solved to locate the bolt \cite{CEJM}. The bolt is located at either 
\begin{equation}
\rho_{bolt}=\frac{1}{4}\,{\frac {1-\sqrt {1+16\,{\lambda}^{2}{L}^{4}-16\,{L}^{2}
\lambda}}{\lambda\,L}} \label{rho1}
\end{equation}
or at
\begin{equation}
\rho_{bolt}=\frac{1}{4}\,{\frac {1+\sqrt {1+16\,{\lambda}^{2}{
L}^{4}-16\,{L}^{2}\lambda}}{\lambda\,L}}\label{rho2}
\end{equation}
provided the quantity under the square
root is non-negative. This last requirement, which automatically guarantees that
$\rho_{bolt}>2L$ as one can check, restricts $L$
:
\begin{equation}
\lambda\,L^{2}\le \left(\frac{1}{2}-\frac{\sqrt{3}}{4}\right)\,\,\,\, (\sim 0.066987298).
\end{equation}
Therefore if $L$ is within this range the bolt is regular, i.e., at
the bolt the metric is the product metric of ${\Bbb E^2}$ and an $S^{2}$ of constant
radius. The
complete metric ($\rho_{bolt} \le \rho <
\infty$) is defined over a manifold having the topology of $\bbbc
P^2$. Readers interested in more details about the global properties of
Taub-Bolt-AdS space are referred to \cite{CEJM,HHP}.

One can now find $M$:
\begin{equation}
M= {\frac {1}{96}}\,{\frac {1 \pm \sqrt {1+16\,{\lambda}^{2}{L}^{4}-16\,\lambda
\,{L}^{2}}\left (32\,{\lambda}^{2}{L}^{4}-8\,\lambda\,{L}^{2}-1\right 
)}{{\lambda}^{2}{L}^{3}}}.\label{M}
\end{equation}
The positive and negative signs correspond to the first and second
values of
$\rho_{bolt}$ above given by Eq. (\ref{rho1}) and Eq.(\ref{rho2}) respectively.
Eq.(\ref{M}) gives $M$ in terms of $L$ and thus reduces the family
from two parameter to one-parameter in the sense that the parameter
takes values freely albeit within a range. However, note that these
two $M$ are not related by a sign and hence they are not just
``orientations'' as in the case of self-dual Taub-NUT-(anti-)de Sitter
family \cite{Akbar1}. This one-parameter family of regular
Taub-Bolt-AdS metrics therefore admits two bifurcated
sub-families. For a given $L$ within the permissible range, there are
therefore two regular Taub-Bolt-AdS metrics.
\section{Infilling Taub-Bolt-AdS geometries}
We first briefly recall the results obtained in \cite{Akbar1} for
self-dual Taub-NUT-(anti-)de Sitter solutions of the same
boundary-value problem. The problem of finding solutions for a given
$S^3$ specified by its two radii $(a,b)$ amounted to solving two
algebraic equations in $\rho$ and $L$:
\begin{equation}
a^{2}-\rho^{2}+L^{2}=0, \label{15}
\end{equation}
and
\begin{equation}
b^{2}(\rho^{2}-L^{2})-4\,{L}^{2}\left (\left (\rho-L\right )^{2}-\frac{1}{3}\,\Lambda\,\left (\rho+3
\,L\right )\left (\rho-L\right )^{3}\right )=0. \label{16}
\end{equation}
This system admits the discrete symmetry $(\rho,L) \leftrightarrow
(-\rho,-L)$.  With simple substitutions that preserve this symmetry,
the above system can be reduced to an algebraic equation of degree
three and hence in general there are three, modulo orientation,
complex-valued infilling solutions for arbitrary $(a,b)$ for which
explicit solutions can be obtained. However, to determine which of
them are real infilling solutions one needs to be more careful. One
finds that there is a unique real, positive-definite self-dual
Taub-NUT-anti-de Sitter solution on the four-ball bounded by a given
Berger-sphere of radii $(a,b)$, if and only if
\begin{equation}
b^2< \frac{1}{3}a^2\,(2a^2|\Lambda|+3). \label{connut}
\end{equation}
If this inequality is not satisfied by the radii $(a,b)$, there is no
real solution at all. It is not, however, a coincidence that the
infilling solution is unique in this case. This is because the set of
ordinary differential equations arising from the Einstein equations
can be converted into first order equations by applying the condition
of self-duality of the Weyl tensor. The boundary data $(a,b)$ then fix
their first derivatives uniquely at the boundary and hence evolution
is unique. The algebraic condition (\ref{connut}) then gives us the
condition on the boundary data for which the two radii of the evolving
nested Berger spheres can go to zero smoothly.

The algebraic system determining whether an $S^3$ boundary can be
filled in classically with a regular Taub-Bolt-AdS geometry similarly
consists of the following two equations in $\rho$ and $L$:
\begin{equation}
a^2-\rho^{2}+L^{2}=0 \label{13}
\end{equation}
and
\begin{equation}
b^2\,\,(\rho^{2}-L^{2})-4\,{L}^{2}\left(\rho^{2}-2M\rho +L^{2}-\lambda( L^{4}+2
L^{2}\rho^{2}-\frac{1}{3}\rho^{4})\right)=0 \label{14}
\end{equation}
where $M$ is given by one of the two choices for corresponding to the
two signs in (\ref{M}). As before $\lambda(=-\Lambda)$ is
positive. However, compared to the self-dual Taub-NUT-(anti-)de Sitter
problem, Eqs.(\ref{13})-(\ref{14}) are slightly unpromising at first
sight. Getting rid of the square roots one would be left with a
polynomial equation of higher degree (degree seven, Eq.(\ref{gp})
below) than we know how to solve analytically. However, a surprising
result follows if one looks carefully:\\ \\ {\bf{Theorem:}} \emph{For
{\bf{any}} boundary data $(a,b)$, where $a$ and $b$ are the radii of
two equal and one unequal axes of a squashed $S^{3}$ respectively,
there exist an odd number of regular Taub-Bolt-AdS solutions bounded
by the $S^{3}$ for either choice of $M$ (and hence, in total, an even
number of regular bolt solutions)}.\\ \\ {\bf{Proof:}} Denote
$(a^2,b^2)$ by $(A,B)$. With the help of the cosmological constant
rescale all quantities so that they are dimensionless:
\begin{equation}
\begin{array}{rcl}
A \lambda,\, B\lambda &\rightarrow& A,\,B \,\,\rm{etc.}\\
\end{array}
\end{equation}
Solve Eq.(\ref{13}) for $\rho$ and substitute into Eq.(\ref{14}). Then, by
suitable squaring, one may obtain, {\it{for either choice of $M$ given by Eq.(\ref{M})}}, the following seventh-degree
equation for $\lambda L^2 \equiv p$, in which no square roots appear:
\begin{equation}
g(p):=a_{7}\, p^{7} +a_{6}\, p^{6}+a_{5}\, p^{5}+a_{4}\, p^{4}+a_{3}\,
p^{3}+a_{2}\, p^{2}+a_{1}\, p+a_{0}=0 \label{gp}
\end{equation}
and where
\\
\begin{equation}
\begin{array}{rcl}
a_{7}\!\!&=\!\!&8192\,A\left (-6\,AB+4\,{A}^{3}+6\,{A}^{2}-1\right )\left (2\,{A}^{2}+
3\,A-3\,B\right ),\\
\\
a_{6}\!\!&=\!\!&-128\,(216\,{A}^{3}-3456\,B{A}^{3}+2304\,{A}^{5}-27+128\,
{A}^{7}-288\,B{A}^{2}-1152\,B{A}^{4}+36\,{A}^{2}\\
&&-36\,BA-1152\,{A}^{3}{
B}^{2}+1840\,{A}^{4}+576\,B{A}^{5}+1728\,{A}^{2}{B}^{2}+960\,{A}^{6}),\\
\\
a_{5}\!\!&=\!\!&4\,(13824\,{A}^{6}+20736\,{A}^{2}{B}^{2}+1728\,{A}^{3}-
4320\,BA+4320\,{A}^{2}+27648\,B{A}^{4}+81+3072\,{A}^{7}\\
&&+1728\,A+256\,{A}^{8}+41472\,B{A}^{5}+1728\,B{A}^{2}+28160\,{A}^{5}+9216\,B{A}^{6}+
20448\,{A}^{4}\\
&&-55296\,{A}^{3}{B}^{2}-39168\,B{A}^{3}),\\
\\
a_{4}\!\!&=\!\!&-8\,A \left( 3240\,AB+10368\,{A}^{4}B-1440\,{A}^{3}-81+5184\,{A}^{3}{B}
^{2}+10752\,{A}^{3}B+72\,{A}^{2}B \right.\\
&&\left. -2176\,{A}^{2}-240\,{A}^{4}-144\,A{B}
^{2}-10368\,{A}^{2}{B}^{2}+3456\,{A}^{4}{B}^{2}+3456\,{A}^{5}B+24\,B \right.\\
&&\left.-
1728\,{A}^{2}{B}^{3}-456\,A+384\,{A}^{6}B \right),\\
\\
a_{3}\!\!&=\!\!&4\,A(-72\,AB+1728\,{A}^{3}{B}^{3}+24\,{A}^{2}-720\,{A}^{3}B+864
\,{A}^{5}{B}^{2}+288\,{A}^{2}{B}^{2}-2160\,{A}^{2}B\\&&+ 5184\,{A}^{4}{B}^{
2}+7776\,{A}^{3}{B}^{2}+270\,A{B}^{2}+36\,B-2592\,{A}^{2}{B}^{3}+45\,A
+28\,{A}^{3} ),\\
\\
a_{2}\!\!&=\!\!&-8\,{A}^{2}\left (-27\,B+648\,{A}^{2}{B}^{3}+9\,AB-135\,A{B}^{2}+216\,
{A}^{3}{B}^{3}+2\,{A}^{3}+18\,A+12\,{A}^{2}\right ),\\
\\
a_{1}\!\!&=\!\!&3\,{A}^{2}B\left (108\,{A}^{2}{B}^{3}+24\,A+8\,{A}^{2}-3\,B\right ),\\
\\
a_{0}\!\!&=\!\!&-9\,A^{3}\,B^{2}.\\
\end{array}
\end{equation}
Recall that regular bolt is possible if $p\, \in
\left(0,\frac{2-\sqrt{3}}{4}\right]$. We want to prove the existence
of an odd number of roots of Eq.(\ref{gp}) within this interval for
arbitrary boundary data $(A,B)$ (and, automatically, for either choice
of $M$).

Since $g(p)$ is of odd degree there will be at least one real
root. Recall that for any function $f(x)$, if $f(a)$
and $f(b)$ have unlike signs then an odd number of roots of $f(x)=0$ lie
between $a$ and $b$ (see, for example, \cite{BC}).
In our case, at the lower limit: 
\begin{equation}
g(0)= -9 A^{3} B^{2}
\end{equation}
which is strictly a negative quantity for any $B \ne 0$. ($B=0$ is naturally
excluded).\\
At the upper limit,
\begin{equation}
\begin{array}{rcl}
g\left(\frac{2-\sqrt{3}}{4}\right)&=&-{\frac {1}{256}}\,\left
(209\,\sqrt {3}-362\right )
 ( -16\,{A}^{4
}-32\,{A}^{3}-32\,{A}^{3}\sqrt {3}+192\,{A}^{3}B+96\,{A}^{3}B\sqrt {3}\\
&-&48\,{A}^{2}-48\,{A}^{2}\sqrt {3}+480\,{A}^{2}B+288\,{A}^{2}B\sqrt {3}
-1008\,{A}^{2}{B}^{2}\\
&-&576\,{A}^{2}{B}^{2}\sqrt {3}+72\,A+60\,A\sqrt {3
}+48\,AB+48\,AB\sqrt {3}-9+12\,\sqrt {3} )^{2}.
\end{array}
\end{equation}
The quantity $-{\frac {1}{256}}\,\left
(209\,\sqrt {3}-362\right )$ is positive and the other quantity is a
square. Therefore $g\left(\frac{2-\sqrt{3}}{4}\right )$ is strictly
non-negative quantity irrespective of the values of $A$ and $B$ and is
zero only when 
\begin{equation}
B={\frac {\frac{2}{3}\,{A}^{2}-\frac{1}{3}\,{A}^{2}\sqrt {3}+\frac{1}{3}\,A\sqrt {3}-\frac{1}{3}\,A-\frac{5}{6} +\frac{1}{2}\,\sqrt {3}\pm \frac{1}{12}\,\sqrt {1+4\,A\sqrt {3}+8\,A}}{A}},\\
\end{equation}
in which case $\left(\frac{2-\sqrt{3}}{4}\right)$ itself is a root of Eq.(\ref{gp}).
So, for \emph{any}
boundary data $(a,b)$, there are \emph{an odd number of roots} within
the interval $\left(0,\frac{2-\sqrt{3}}{4}\right]$, and hence an even number of regular bolt solutions (for the
two choices of $M$). QED

Therefore there will at least be two Taub-Bolt-AdS infillings for any
biaxial $S^{3}$ boundary. This is therefore in sharp contrast with the
self-dual Taub-NUT-AdS case.
\section{Region of physical interest: Isotropy and low anisotropy }
The polynomial equation (\ref{gp}) is of degree seven with arbitrary
coefficients determined by the values of the two radii of the Berger
sphere. From Galois theory we know that general polynomial equations
of degree five or above are not solvable by radicals. However, they
can be solved by higher order hypergeometric functions of several
variables \cite{Birkeland}. For our purpose pursuing this line would
not be illuminating. The remaining approach is to solve Eq.(\ref{gp})
numerically for $p$ and check individually whether the solutions fall within
the range $\left( 0,\frac{2-\sqrt{3}}{4}\right]$. These steps are
necessary should one seek the infilling geometries and their actions
as functions of the boundary data and are left for future work. In
this paper we instead count the number of possible infilling solutions
without solving Eq.(\ref{gp}) numerically and see how they vary as the
boundary data is varied. Indeed there are elegant ways of doing it
without having to find explicit numerical solutions as will be
described below.

From a physical point of view we are more interested in the
qualitative behaviour when the squashing $a/b$ is not too high, i.e.,
when $a$ and $b$ are roughly of the same order of magnitude, and they
are not too small. As we will see below, it is possible to make
general statements on the possible number of infilling Taub-Bolt-AdS
solution in this case. However, there are more surprises for large
anisotropy which will be described in Section 5.

To proceed further we would need to recall Fourier's
theorem \cite{BC}: \emph{If $f(x)$ is a polynomial of degree $n$ and
$f_{1},f_{2},...f_{n}$ are its successive derivatives, the number
of real roots $R$ which lie between two real numbers $p$ and $q$
($p<q$) are such that $R\le N-N'$,
where $N$ and $N'$ ($N\ge N'$) respectively denote the number of
changes of sign in the sequence $f_{1},f_{2},...f_{n}$, when $x=p$ and
when $x=q$. Also $((N-N')-R)$ is an even number or zero.}

A formal way of proceeding would be to compute the derivatives of
$g(p)$, $g'(p)$, $g''(p)$ etc. and evaluate them at zero and at
$\left(\frac{2-\sqrt{3}}{4}\right)$ and check their values setting $a$ and $b$
roughly equal. However, for large radii and small
anisotropy, it is sufficient and more illuminating to set
$A=B$ in Eq.(\ref{gp}). The justification for setting them equal while
considering anisotropy will
be given below. With this substitution the coefficients of
Eq.(\ref{gp}) simplify:
\begin{equation}
\begin{array}{rcl}
a_{7}\!\!&=\!\!&16384B^3(4B^3-1),\\
\\
a_{6}\!\!&=\!\!&-128(128\,B^7+1536\,B^6+112\,B^4-72\,B^3-27),\\
\\
a_{5}\!\!&=\!\!&4(256\,{B}^{8}+12288\,{B}^{7}+55296\,{B}^{6}+512\,{B}^{5}+2016\,{B}^{4}+
3456\,{B}^{3}+1728\,B+81),\\
\\
a_{4}\!\!&=\!\!&-8\,B(384\,{B}^{7}+6912\,{B}^{6}+13824\,{B}^{5}+144\,{B}^{4}-1512\,{B}^{3}+
1064\,{B}^{2}-432\,B-81),\\
\\
a_{3}\!\!&=\!\!&4\,B^2\left (864\,{B}^{6}+6912\,{B}^{5}+5184\,{B}^{4}-432\,{B}^{3}-1862
\,{B}^{2}-48\,{B}+81\right ),\\
\\
a_{2}\!\!&=\!\!&-8\,{B}^{3}\left (216\,{B}^{5}+648\,{B}^{4}-133\,{B}^{2}+21\,{B}-9\right ),\\
\\
a_{1}\!\!&=\!\!&  3\,{B}^{4}\left (108\,{B}^{4}+8\,{B}+21\right ),       \\
\\
a_{0}\!\!&=\!\!&-9\,B^{5}.\\ \label{iso}
\end{array}
\end{equation}
Note that the coefficients have been written in decreasing powers of
$B$. It is now straightforward to check that, for
$B$ not too small, the signs of 
coefficients of different powers of $p$ are insensitive to the exact
value of $B$ by observing that $a_{0}$, $a_{1}$, $a_{5}$ are positive
definite for any $B>0$ and that $a_{7}$, $a_{6}$, $a_{4}$, $a_{3}$,
$a_{2}$ are positive definite for values of $B$ (approximately)
greater than $0.6299605250$, $0.5123378480$, $0.4031183975$,
$0.4565392083$ and
$0.4247313485$ respectively. Therefore if $B>0.6299605250$ the signs
of the coefficients $a_{i}$ will not change. 

We now consider small anisotropy. Note that in any $a_{i}$ in Eq.(
\ref{iso}) above the highest two powers of $B$ always occur with the
same sign and hence the results will be unchanged if $A$ and $B$ are
not equal and instead are roughly of the same order since the degree
$(n+m)$ of the two highest powers (now of the form $A^n B^m$) will
dominate. Also note that in each of the coefficients $a_{i}$ in
(\ref{iso}) above the coefficients of lower powers of $B$ which have
opposite signs to the highest two powers are not so large as to change
the sign of $a_{i}$ so long as $A$ and $B$ are reasonably large.

It is now easy to see that coefficients of $g(p)$ occur with
alternating signs. This immediately gives the sign of the successive
derivatives of $g(p)$ evaluated at zero, thus leaving us with the task
of evaluating derivatives at $\left(\frac{2-\sqrt{3}}{4}\right)$ in
order to apply Fourier's theorem.

In evaluating the derivatives at $\left(\frac{2-\sqrt{3}}{4}\right)$
(henceforth denoted by $q$ for brevity) one should check whether
arguments similar to those made above can be applied to this case.
The various derivatives of $g(p)$ in this case are:
\begin{equation}
\begin{array}{rcl}
g_{1}\!\!&=\!\!& 135.4256260(B+ 0.4644486459)(B+0.2555353665)(B+ 0.1313663488)(B+
 0.06555587950)\\
&& (B- 1.141398520)(B-
 1.993301852)({B}^{2}- 0.6177105744B+ 0.1923992276),\\
\\
g_{2}\!\!&=\!\!&- 2226.215022(B+ 1.268856328)(B+
 1.096797061)(B+ 0.07953842687)(B-
 0.4841144260)\\
&&({B}^{2}+ 0.4225153212B+ 0.04767173654)({B}^{2}- 1.508758093\,B+
0.7095903619),\\
\\
g_{3}\!\!&=\!\!&16072.86008(B+ 5.568090955)(B+
 0.5367072895)(B+ 0.2196287557)(B+
 0.1046514573)\\
&& ({B}^{2}+ 0.3588828520B+ 0.5018604161)({B}^{2}- 1.211388061B+ 0.3824418783),\\
\\
g_{4}\!\!&=\!\!&- 65496.60082(B+ 13.28384331)(B+
 0.3400846060)(B+ 0.1452308623)\\
&&(B-
 0.6346052509)({B}^{2}+ 1.920945736B+ 1.048214873)({B}^{2}-
0.4217016508B+ 0.2864195974),\\
\\
g_{5}\!\!&=\!\!&122880.0(B+ 37.73129542)(B+ 3.820066961)(B+
0.6000369611)(B+ 0.2248301376),\\
&&
({B}^{2}+ 0.08583004940B+ 0.2599653669)({B}^{2}-
 0.8928401318B+ 0.3309398116),
\\
\\
g_{6}\!\!&=\!\!&- 11796480.0(B+ 10.13296728)(B+
 0.4974325158)(B- 0.5008149571),\\
&&({B}^{2}+
 0.4893229700B+ 0.2971783280)({B}^{2}- 0.4945521540B+
 0.2811826749),\\
\\
g_{7}\!\!&=\!\!&82575360\,\left (-1+4\,{B}^{3}\right ){B}^{3}.\label{420}
\end{array}
\end{equation}
Note that for any $B>1.993301852$, which is true by assumption, the signs
of the derivatives do not change as one goes to higher values of $B$. We now list the results in tabular form
below:
 \begin{center}
  \begin{tabular}[c]{c |c c}
    
     & $0$ & $(\frac{1}{2}-\frac{\sqrt{3}}{4})$ \\ \hline
    
    $g(p)$ & $-$ & $+$ \\ 
     $g_{1}(p)$& $+$ & $+$ \\ 
    $g_{2}(p)$ & $-$ & $-$ \\ 
    $g_{3}(p)$ & $+$ & $+$  \\ 
    $g_{4}(p)$ & $-$ & $-$ \\ 
    $g_{5}(p)$ & $+$ & $+$  \\
    $g_{6}(p)$ & $-$ & $-$ \\ 
    $g_{7}(p)$ & $+$ & $+$  \\ \hline
   sign  && \\
   changes & $7$ & $6$
  \end{tabular}
\end{center}
The number of roots is the difference between the changes of
sign which is one. Therefore for low squashing and not-too-small radius
there will precisely be two infilling 4-geometries (corresponding to
two choices of $M$) which contain regular bolts inside. 
\subsection{Round $S^3$}
We have already mentioned that an $SO(4)$-invariant $S^3$ can always be
filled in uniquely by parts of $H^4$ and with a unique self-dual
Taub-NUT-AdS solution. One therefore is naturally interested
in how a round $S^3$ can be filled in with 
Taub-Bolt-AdS metrics. \\
\\
{\bf{Lemma:}} {\it{An $SO(4)$-invariant $S^3$ of arbitrary radius
admits two Taub-Bolt-AdS infillings (for the two
choices of $M$).}}\\
\\
{\bf{Proof:}} 
The above analysis has already
established that an $SO(4)$-invariant $S^3$ boundary can be 
filled in with two Taub-Bolt-AdS solutions assuming
$B(=A)>1.993301852$. We noted that the signs of the
various derivatives of $g(p)$ at the two limits of $p$ are not
strictly the same for $B<1.993301852$. It is not difficult to check that 
within the interval $(0, 1.993301852)$ the sign changes of
various $g_{n}(0)$ are
always compensated by sign changes of $g_{n}\left(\frac{2-\sqrt{3}}{4}\right)$ and thus the number of
roots within the interval remains unchanged. It has been done numerically. We
have counted the number of roots for low values of the radius with a
higher degree of precision than in Eq.(\ref{420}) which is shown in Fig. 1. Therefore a
round sphere has two Taub-Bolt-AdS metrics irrespective
of its radius. 
\begin{figure}[!h]
  \begin{center}
    \leavevmode
    \vbox {
      \includegraphics[width=8cm,height=6cm]{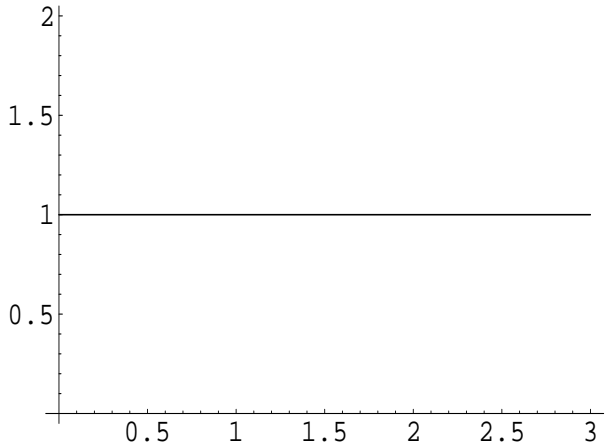}
      \vskip 3mm
      }
    \caption{The number of Taub-Bolt-AdS infilling solutions (for either
    choice of $M$) is one for an $SO(4)$-invariant $S^3$
    irrespective of it radius.} 
      \end{center}
\end{figure}
\section{Anisotropy and number of infilling Taub-Bolt-AdS geometries}
As mentioned in the previous section, it is only possible to make
general statements for isotropic or slightly anisotropic
cases. However, one is naturally interested in what happens when the
Berger sphere is made more anisotropic. Does the number of solutions
change or remain the same? Note that the values, and not just the
ratio, of the boundary data $a$, $b$ are important. This is because
the presence of a cosmological constant in the interior of the Berger
sphere introduces a scale. Therefore the solutions will change if $a$
and $b$ are varied while their ratio is kept constant unlike the case
of zero cosmological constant.

In the isotropic limit the problem can be reduced to a one dimensional
one as we have seen in the previous section. We have counted the
number of roots in a given interval as we varied the ``parameter''
$B$. The same procedure can be adopted for counting the number of
roots as $A$ and $B$ are varied. Recall that the number of roots is an
odd number and hence as $A$ and $B$ are varied the number is expected
to jump in steps of two catastrophically or remain unity as in the
isotropic limit. The results are plotted in Figs. 2-4 confirming the
catastrophic jumps of the solutions. However, slightly to our
surprise, we find that the number of roots of Eq.(\ref{gp}) can be as
high as five. This is somewhat unexpected given that we are seeking
solutions within a very narrow prescribed range of $L$.
\begin{figure}[!h]
{\includegraphics[width=0.45\textwidth]{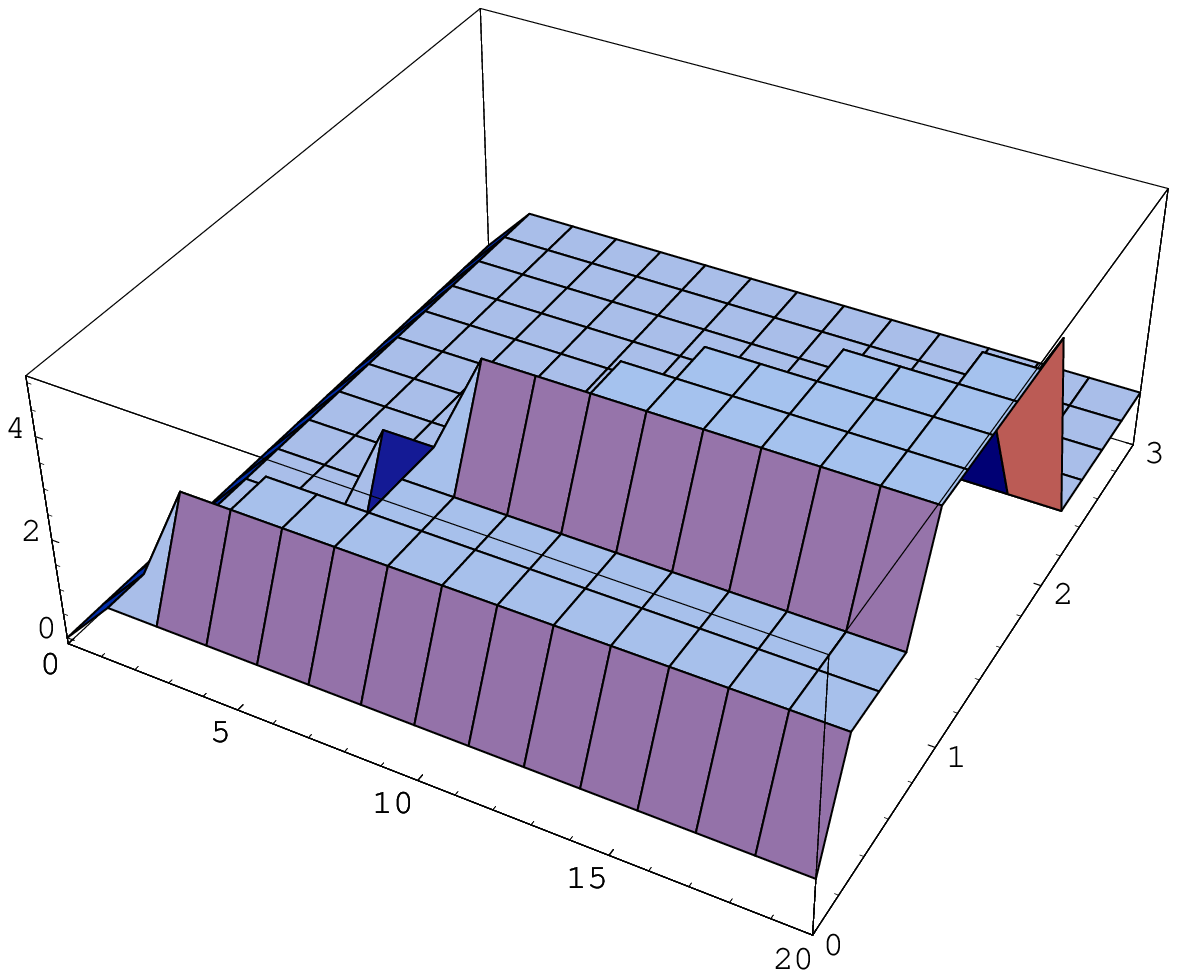}}
\hspace*{1.2cm} 
{\includegraphics[width=0.45\textwidth]{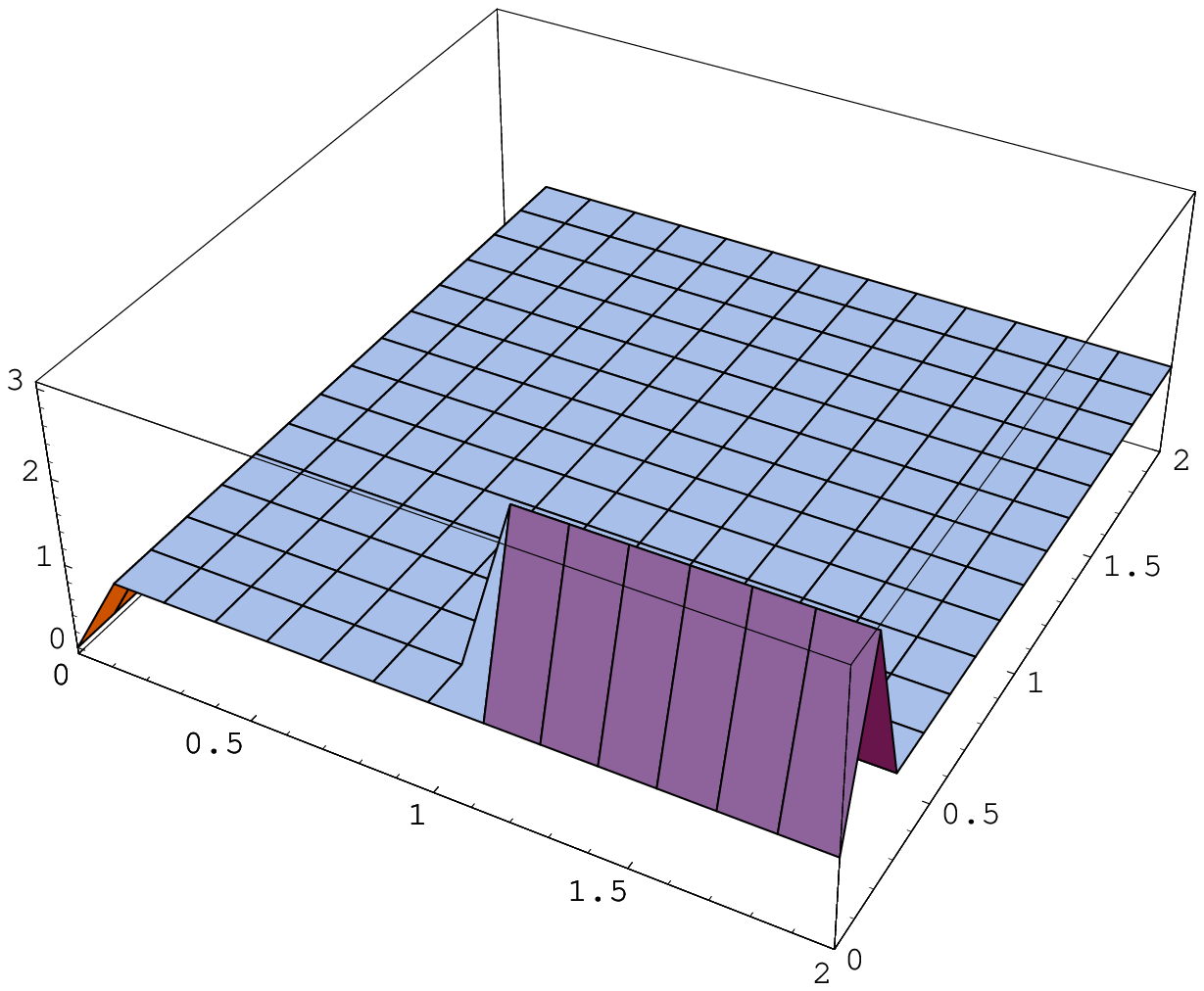}}
\caption{The number of infilling solutions changes catastrophically
and can be as many as five.}
\end{figure}
\begin{figure}[htb]
{\includegraphics[width=0.45\textwidth]{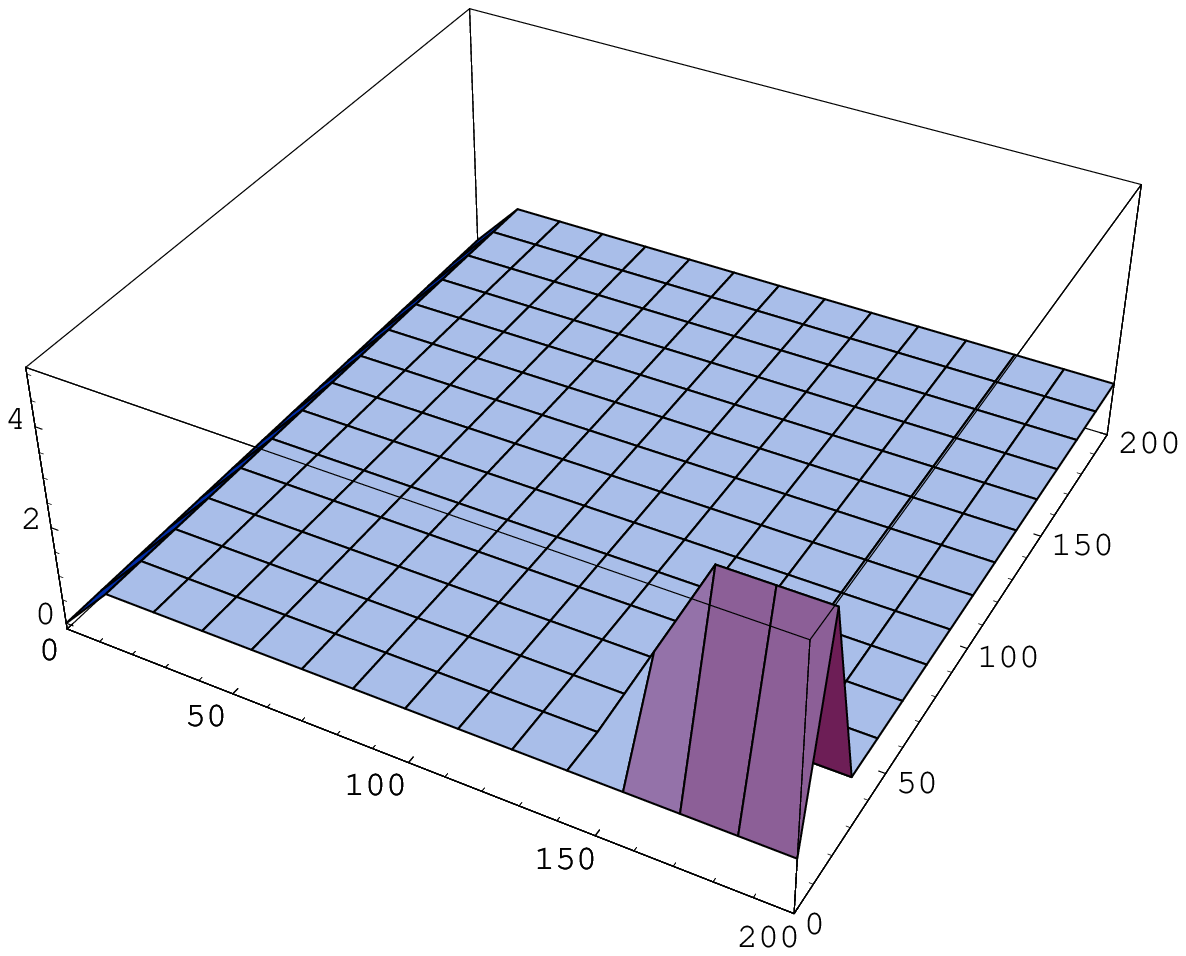}}
\hspace*{1.2cm} 
{\includegraphics[width=0.45\textwidth]{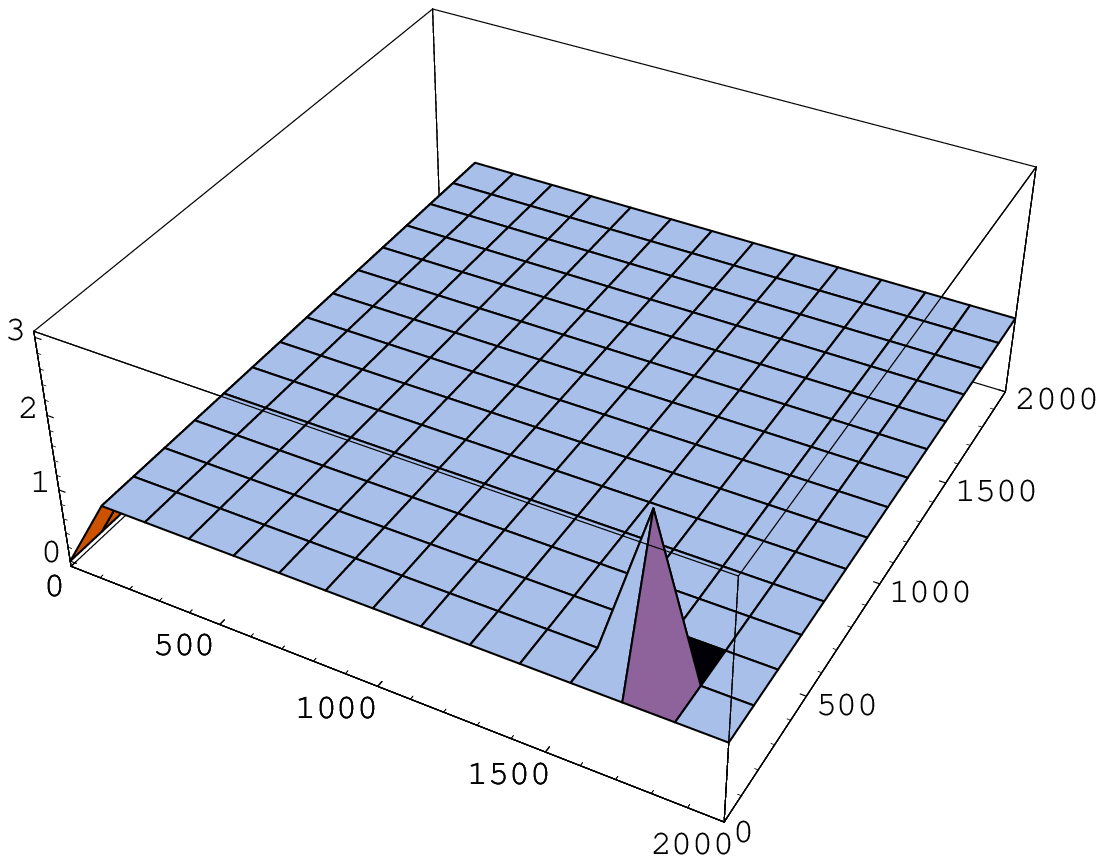}}
\caption{$A$ and $B$ of the order of 100-1000. The solutions can be as
many as five even when $A$ and $B$ are not too small. Note that there
is an averaging effect.}
\end{figure}
\begin{figure}[!h]
  \begin{center}
    \leavevmode
    \vbox {
      \includegraphics[width=9.5cm,height=8.3cm]{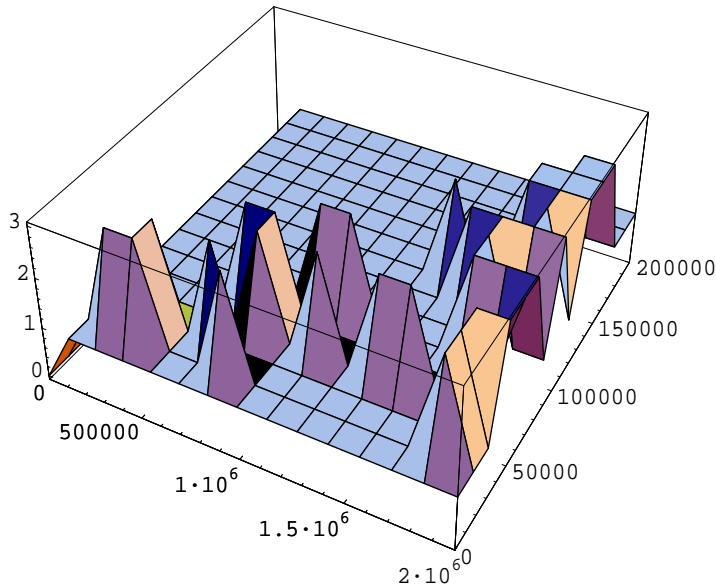}
      \vskip 3mm
      }
\caption{$A$ and $B$ very large: for $A \sim 10 B$ multiple roots
appear; the nature of the averaging effect in the figure means that higher number of
solutions are rarer. Only in very small regions five
roots will appear. Note that close to the isotropic limit the number of roots is
clearly one as in Fig 3.}
      \end{center}
\end{figure}
\section{Conclusion}
In this paper we have studied the filling in problem for an axially
symmetric $S^3$ boundary with regular Taub-Bolt-AdS
metrics\footnote{The filling in problem for a conformal boundary of
$S^{3}$ topology for selfdual spaces have been studied by Pedersen
\cite{ped}, Hitchin \cite{Hit}, Tod \cite{Tod}. Also the Dirichlet
problem for general conformal boundaries has been studied in
connection with AdS/CFT correspondence (see \cite{Skend1,Skend2} and
references therein).}. The same boundary-value problem with self-dual
Taub-NUT-AdS was studied in \cite{Akbar1} and it was found that an
$S^3$-boundary can be filled in with a unique or no real self-dual
Taub-NUT-anti-de Sitter metric depending on whether the two radii of
the $S^3$ satisfy the inequality (\ref{connut}) or not\footnote{Here
we mean the self-dual Taub-NUT-(anti-)de Sitter metrics in which the
parameter is non-zero and finite for which the metric is well-defined
in its usual form. As explained in \cite{Akbar1} other solutions exist
as special limits to this metric which should be included if one
considers the larger class of self-dual Bianchi-IX Einstein infilling
solutions.}. The Taub-Bolt-AdS metrics contain two-dimensional
fixed-point sets at the centre and can be extended globally over a
manifold with the topology of $\bbbc P^2$.  The condition of
regularity of the bolt constrains the free parameter $L$ to take
values within a very narrow range on the real line. Also, as we have
noted, there is a two-fold degeneracy in the family in that the other
parameter $M$ is a 1-2 function of $L$.

The restriction on the permissible values of $L$ is rather stringent
especially compared to the self-dual Taub-NUT-(anti-)de Sitter and the
Taub-Bolt-de Sitter families for all of which $L$ can take values
freely without compromising regularity. In this case therefore one
would naturally expect that such an {{\it a priori}} restriction on
$L$ would put some limits on the two radii for the Berger sphere to
qualify for having regular Taub-Bolt-AdS infillings. However, slightly
contrary to these intuitions, we found that there will always be at
least one regular Taub-Bolt-AdS infilling solution (for each of the
two sub-families) irrespective of the magnitude of the two radii of
the Berger sphere. This is comparable to filling a round $S^3$ with
part of $H^4$ which is possible irrespective of the radius of the
$S^3$ as we have mentioned earlier. We also found that in the isotropy
and low-anisotropy limits (and if the radii are not too small), the
solution is unique for either of two branches. For perfect isotropy
this holds for small radii as well. Furthermore, we showed that the
number of such infilling solutions varies abruptly from one to five in
steps of two and hence they form a catastrophe-structure in which $a$
and $b$ play the role of control parameters (see, for example,
\cite{PS}). Compare it with Schwarzschild-AdS in four dimensions
\cite{BCM} in which case the boundary is the trivial bundle $S^1\times
S^2$ and can admit only two black hole solutions or no solutions
depending on the two radii. We conjecture that Taub-Bolt-AdS metric
(and those with other values of the integer $k$) and their higher
dimensional generalisations are the only Einstein metrics for which
such a large number of regular infilling solutions can occur. In the
standard examples in the literature with Einstein metrics (including
those with $\Lambda=0$) this number, i.e., the number of infilling
solutions for a given type of metrics does not exceed two. This paper
provides the first example in which the number of infilling solutions
exceeds two and can be as large as five, i.e., ten for the two
branches put together, depending on the boundary data.
\subsection*{Acknowledgements}
The author would like to thank Peter D'Eath and Gary Gibbons for
helpful discussions during the preparation of the manuscript. The
author was supported by awards from the Cambridge Commonwealth Trust
and the Overseas Research Scheme and by DAMTP.


\begin{thebibliography}{99}
\bibitem{Akbar1}
M.~M.~Akbar and P.~D.~D'Eath,
``Classical Boundary-value Problem in Riemannian Quantum Gravity and
Self-dual Taub-NUT-(anti-)de Sitter Geometries,'' 
Nucl.\ Phys.\ B {\bf 648} (2003) 397
[arXiv:gr-qc/0202073].

\bibitem{BC} S.~Barnard, and J.~M.~Child (1936). \emph{Higher Algebra}
(Macmillan, India).

\bibitem{Birkeland} 
R.~Birkeland,
``\"{U}ber die Aufl\"{o}sung algebraischer Gleichungen durch
hypergeometrische Funktionen,''
Math. Zeitschrift 26 (1927) 565-578.

\bibitem{BCM}
J.~D.~Brown, J.~Creighton and R.~B.~Mann,
``Temperature, Energy and Heat Capacity of Asymptotically Anti-De Sitter Black Holes,''
Phys.\ Rev.\ D {\bf 50} (1994) 6394
[arXiv:gr-qc/9405007].

\bibitem{Carter}
B.~Carter,
``Hamilton-Jacobi And Schrodinger Separable Solutions Of Einstein's Equations,''
Commun.\ Math.\ Phys.\  {\bf 10} (1968) 280.
\bibitem{CEJM}
A.~Chamblin, R.~Emparan, C.~V.~Johnson and R.~C.~Myers,
``Large N phases, gravitational instantons and the nuts and bolts of AdS  holography,''
Phys.\ Rev.\ D {\bf 59} (1999) 064010
[arXiv:hep-th/9808177].

\bibitem{Clarkson1}
R.~Clarkson, L.~Fatibene and R.~B.~Mann,
``Thermodynamics of (d+1)-dimensional NUT-charged AdS spacetimes,''
arXiv:hep-th/0210280.

\bibitem{EGH}
T.~Eguchi, P.~B.~Gilkey and A.~J.~Hanson,
``Gravitation, gauge theories and differential geometry,''
Phys.\ Rep.\  {\bf 66} (1980) 213.

\bibitem{Emparan1}
R.~Emparan, C.~V.~Johnson and R.~C.~Myers,
``Surface terms as counterterms in the AdS/CFT correspondence,''
Phys.\ Rev.\ D {\bf 60} (1999) 104001
[arXiv:hep-th/9903238].

\bibitem{Ghezelbash1}
A.~M.~Ghezelbash and R.~B.~Mann,
``Nutty bubbles,''
JHEP {\bf 0209} (2002) 045
[arXiv:hep-th/0207123].

\bibitem{GP1}
G.~W.~Gibbons and C.~N.~Pope,
``The Positive Action Conjecture and Asymptotically Euclidean Metrics in Quantum Gravity,''
Commun.\ Math.\ Phys.\  {\bf 66} (1979) 267.

\bibitem{HHP}
S.~W.~Hawking, C.~J.~Hunter and D.~N.~Page,
``Nut charge, anti-de Sitter space and entropy,''
Phys.\ Rev.\ D {\bf 59} (1999) 044033
[arXiv:hep-th/9809035].

\bibitem{HP}
S.~W.~Hawking and D.~N.~Page,
``Thermodynamics of Black Holes in Anti-De Sitter Space,''
Commun.\ Math.\ Phys.\  {\bf 87} (1983) 577.

\bibitem{Hit}
N.~J.~Hitchin,
``Twistor Spaces, Einstein Metrics and Isomonodromic Deformations,'' 
J.\ Differential\ Geom. {\bf 42} (1995), no. 1, 30--112.

\bibitem{GH} G.~W.~Gibbons and S.~W.~Hawking, ``Classification Of
Gravitational Instanton Symmetries,'' Commun.\ Math.\ Phys.\ {\bf 66}
(1979) 291.

\bibitem{Louko}
L.~G.~Jensen, J.~Louko and P.~J.~Ruback,
``Biaxial Bianchi Type IX Quantum Cosmology,''
Nucl.\ Phys.\ B {\bf 351} (1991) 662.

\bibitem{Maldacena}
J.~M.~Maldacena,
``The Large N Limit of Superconformal Field Theories and Supergravity,''
Adv.\ Theor.\ Math.\ Phys.\  {\bf 2} (1998) 231
[Int.\ J.\ Theor.\ Phys.\  {\bf 38} (1999) 1113]
[arXiv:hep-th/9711200].

\bibitem{Mann1}
R.~B.~Mann,
``Entropy of rotating Misner string spacetimes,''
Phys.\ Rev.\ D {\bf 61} (2000) 084013
[arXiv:hep-th/9904148].

\bibitem{sak} T. Sakai, ``Cut loci of Berger's spheres'', Hokkaido
Math. J. {\bf{10}} (1981), no. 1, 143-155.

\bibitem{Skend1}
K.~Skenderis,
``Asymptotically Anti-de Sitter Spacetimes and Their Stress Energy  Tensor,''
Int.\ J.\ Mod.\ Phys.\ A {\bf 16} (2001) 740
[arXiv:hep-th/0010138].


\bibitem{Skend2}
K.~Skenderis,
``Lecture Notes on Holographic Renormalization,''
Class.\ Quant.\ Grav.\  {\bf 19} (2002) 5849
[arXiv:hep-th/0209067].

\bibitem{Page}
D.~N.~Page,
``Taub-Nut Instanton with an Horizon,''
Phys.\ Lett.\ B {\bf 78} (1978) 249.

\bibitem{ped} H. Pedersen, ``Einstein metrics, spinning top motions
and monopoles,''
Math. Ann. {\bf 274} (1986) 35.

\bibitem{PS} T.~Poston and I.~Stewart (1978). \emph{Catastrophe Theory
and its Applications} (Pitman, London).

\bibitem{Tod}
K.~P.~Tod,
``Self-dual Einstein metrics from the Painlev\'{e} VI equation,'' 
Phys.\ Lett.\ A {\bf 190} (1994) 221.
\end{thebibliography}
\end{document}